
\magnification=1200
\parindent=0pt
{\it Waves and Particles in Light and Matter, pp. 477-480,
     eds.: A.Garuccio and A. van der Merwe
     (Plenum, New York, 1994); also qr-qc/9410036.}
\parindent=20pt
\bigskip
\centerline{\bf Quantum Particle As Seen In Light Scattering}
\centerline{Lajos Di\'osi}
\centerline{KFKI Research Institute for Particle and Nuclear Physics}
\centerline{H-1525 Budapest 114, POB 49, Hungary}
\centerline{E-mail: diosi@rmki.kfki.hu}
\bigskip
Abstract
\bigskip

A possible mathematical model
has been proposed for motion of illuminated quantum particles seen by
eyes or similar devices  mapping the scattered light.
\bigskip\bigskip
Key words: wavefunction localization, Ito-stochastic equation
\bigskip\bigskip
Spatial motion of free quantum particles is described by the
Schr\"odinger wave equation:
$$
{d\over dt}\psi(x) = {i\over2m}\Delta\psi(x).
\eqno(1)$$
Typical solutions of this equation show an unlimited growths of the
wave packet width $\sigma$.

If the particle is illuminated then the
scattered light will show the trajectory of the quantum particle. Of
course, the free particle Schr\"odinger equation (1) will no longer
be valid. If a single photon of incoming wavenumber $k_i$ has scattered
with the final wavenumber $k_f$ then the scatterer's wavefunction
obtains a unitary factor:
$$
\psi_i(x)\rightarrow\psi_f(x\vert\theta)\equiv
        exp\left(ik_{fi}x\right)\psi_i(x)
\eqno(2a)$$
where the scattering angle $\theta$ between $k_i$ and $k_f$, resp.,
is distributed according to the modulus
square of the scattering amplitude $f$:
$$
p(\theta)={1\over4\pi}\vert f(\theta)\vert^2.
\eqno(2b)$$
The processes (2ab) interrupts the otherwise deterministic evolution (1)
of the wavefunction.

We shall assume a certain
diffuse light of wavenumber $k$.
Then the scattering processes will occur
randomly at a given rate denoted by $\nu$. To get an easy insight into
the resulting stochastic process, we shall assume that
$\lambda=2\pi/k$ is much
bigger than $\sigma$ and, furthermore, that the repetition frequency $\nu$
of scatterings is big as compared to the time scale of the free dynamics
(1) of the particle. The following approximate equation will then
describe the dynamics of the illuminated quantum particle:
$$
{d\over dt}\psi(x) = {i\over2m}\Delta\psi(x)-iF(t)x\psi(x)
\eqno(3)$$
where $F(t)$ is a certain stationary white noise
of correlation $\gamma\delta(t)$. The dispersion takes the form
$$
\gamma=\nu k^2\int4sin^2(\theta/2)p(\theta){d\Omega\over4\pi}.
\eqno(4)$$

The eq.(3) is "almost" an ordinary Schr\"odinger equation. The
particle's motion is influenced by the random force $F(t)$. This
force makes fuzzy phases to the wavepacket but does not prevent its
delocalization. Hence, the linear stochastic eq.(3) is not suitable
to represent the experienced tarjectory of the quantum particle.

What is wrong with our attempt? Where have we lost the well shaped
positions and trajectories of our particle clearly {\it seen} when
illuminated?

We have to go back to the elementary scattering process (2ab). Observe
that the jumps (2a) of the wavefunction assumes {\it a particular}
classification of the photon states. Actually, the final states (as well
as the initial ones) have been classified according to their momenta.
Obviously, trajectories can not be observed via the scattering {\it angles}
of the photons. They are only observed by identifying the position
where the scattered light has emerged. One has to think of a lense
inserted on the path of each scattered photon, making an optical map
of the scatterer particle.

Let us present the mathematical model of the above set up. Introduce
the special Fourier transform of the scattering amplitude:
$$
\ell(x\vert\xi)\equiv
{1\over4\pi}\int e^{ik_{fi}(x-\xi)}\vert f(\theta)\vert^2d\Omega.
\eqno(5)$$
The influence of a single scattering process on the particle's wavefunction
can thus be written as:
$$
\psi_i(x)\rightarrow\psi_f(x\vert\xi)\equiv
        {\cal N}^{-1}(\xi)\ell(x\vert\xi)\psi_i(x).
\eqno(6a)$$
This jump differs very much from the previous one (2a). The jump (6a) is
nonlinear and {\it localizes} the wave
function in the neighbourhood of the random position $\xi$. The
probablity distribution of $\xi$ is simply related to the normalization
factor of the "reduced" wavefunction:
$$
p(\xi)={\cal N}^2(\xi).
\eqno(6b)$$

In the same approximation that we assumed for the process (2ab), the
nonlinear process (6ab) leads to the following nonlinear counterpart of the
phenomenological eq.(3):
$$
{d\over dt}\psi(x) = {i\over2m}\Delta\psi(x)-F(t)(x-<x>)\psi(x)
                                            -{\gamma\over2}(x-<x>)^2\psi(x).
\eqno(7)$$
The derivation of this
Ito-stochastic differential equation can be learnt from Ref.[1].

Let us make a formal comparison of eq.(7) to the "almost" ordinary
Schr\"odinger eq.(3). The main difference on the rhs is the lack of the
factor $i$ in the second term as if the random force had become pure
imaginary. (The third term then comes in just to restore normalization of
the wave function.) To the authors knowledge,
this {\it imaginary} random force has no direct
physical interpretation but that it has come out from eqs.(1) and (6ab)
in the given approximation.

In a search for stochastic equation
of state reduction Gisin [2] had, in fact, discovered the related couple
of eqs.(3) and (7) and showed that eq.(3) did not reduce while eq.(7) did.
{\it Mutatis mutandis}: we mentioned above that eq.(3) possesses no
trajectory-like solutions while, by now not very surprisingly,
the nonlinear counterpart (7) has shown to have them.

In Ref.[3] we have calculated the stationary solutions of the eq.(7).
These solutions are squeezed Gaussian wave packets:
$$
\psi(x)=\left(2\pi\sigma_\infty^2\right)^{-3/4}
        exp\left(ix<p>-{1-i\over4\sigma_\infty^2}(x-<x>)^2\right)
\eqno(8)$$
of stationary width $\sigma_\infty=(2/m\gamma)^{1/4}$. The
centrum of the wave packet follows the classical straightline
trajectories with $d<x>/dt=<p>/m=const.$
apart from a tiny random walk around it. Exact results [3,4] mimic
as if the particle momentum $<p>$ were influenced by the {\it real}
random force $F(t)$ which was not the case at the level of the
Schr\"odinger eq.(7). Nevertheless, the stationary random walk
of the averaged momentum $<p>$ turns out to satisfy the Newton
equation
$$
{d\over dt}<p>=F(t).
\eqno(9a)$$
The quantum average of the position satisfies the following
equation:
$$
m{d\over dt}<x>=<p>+2m\sigma_\infty^2F(t)
\eqno(9b)$$
containing the anomalous second term on its rhs.
It would, of course, be important
to discuss how the quality of trajectories depends on the intensity
and on the spectrum of the light.

In summary, we claim that the processes (6ab) and, especially,
the fenomenological eq.(7) have turned out to be
a suitable mathematical model for the seen/detected motion of
an illuminated quantum particle. It is, furthermore, worthwhile
to mention a few further applications [5],[6],[7],[8]
of nonlinear stochastic
Schrodinger equations, related to  the quantum
measurement problem.
\bigskip

This work was supported by the Hungarian Scientific Research Fund
under Grant No 1822/1991.
\bigskip

References
\bigskip
\parindent=0pt

[1] L.Di\'osi, Phys.Lett. {\bf 129A}, 419 (1988)

[2] N.Gisin, Phys.Rev.Lett. {\bf 52}, 1657 (1984)

[3] L.Di\'osi, Phys.Lett. {\bf 132A}, 233 (1988)

[4] D.Gatarek and N.Gisin, J.Math.Phys. {\bf 32}, 2152 (1991)

[5] N.Gisin, Helv.Phys.Acta, {\bf 63}, 929 (1989)

[6] L.Di\'osi, Phys.Rev. {\bf A40}, 1165 (1989)

[7] G.C.Ghirardi,P.Pearle and A.Rimini, Phys.Rev. {\bf A42}, 78 (1990)

[8] A.Barchielli and V.P.Belavkin, J.Phys. {\bf A24}, 1495 (1991)

\vfill
\end